\documentclass[%
 reprint,
superscriptaddress,
 amsmath,amssymb,
 aps,
pra,
prb,
rmp,
prstab,
prstper,
floatfix,
]{revtex4-2}
\usepackage{romannum}
\usepackage{graphicx}
\usepackage{dcolumn}
\usepackage{bm}
\usepackage[hidelinks,breaklinks,colorlinks=true,linkcolor=blue,citecolor=blue,urlcolor=blue]{hyperref}
\makeatletter
\let\old@makecaption=\@makecaption
\usepackage{subcaption}
\let\@makecaption=\old@makecaption
\makeatother

\begin{document}

\title{Self-Consistent Description of Vapor-Liquid Interface in Ionic Fluids}
\date{\today}
\author{Nikhil R. Agrawal}
\affiliation{ 
Department of Chemical and Biomolecular Engineering, University of California, Berkeley, California 94720, USA
}%
\author{Rui Wang}
\email{email: ruiwang325@berkeley.edu}
\affiliation{ 
Department of Chemical and Biomolecular Engineering, University of California, Berkeley, California 94720, USA
}
\affiliation{ 
Materials Sciences Division, Lawrence Berkeley National Laboratory, Berkeley, 94720, California, USA
}

\begin{abstract}
Inhomogeneity of ion correlation widely exists in many physicochemical, soft matter, and biological systems. Here, we apply the modified Gaussian renormalized fluctuation theory to study the classic example of the vapor-liquid interface of ionic fluids. The ion correlation is decomposed into a short-range contribution associated with the local electrostatic environment and a long-range contribution accounting for the spatially varying ionic strength and dielectric permittivity. For symmetric salt, both the coexistence curve and the interfacial tension predicted by our theory are in quantitative agreement with simulation data reported in the literature. Furthermore, we provide the first theoretical prediction of interfacial structure for asymmetric salt, highlighting the importance of capturing local charge separation. 
\end{abstract}

\maketitle

One outstanding challenge in the physical chemistry of charged systems is to capture the inhomogeneity in electrostatic correlation resulting from spatially varying ionic strength or dielectric permittivity. A solution to this problem is critical to explain many important phenomena such as salt concentration and specific ion effect on the surface tension\cite{Jungwirth2006SpecificInterface,Levin2009IonsMystery}, stability of colloidal and protein solutions\cite{Grosberg2002Colloquium:Systems, Zhang2008ReentrantCounterions,Caccamo_2000}, electrokinetic flow in nano-devices\cite{Schoch2008TransportNanofluidics,VanDerHeyden2006ChargeCurrents}, ion transport in energy storage materials\cite{Hallinan2013PolymerElectrolytes}, and interfacial properties of charged macromolecules\cite{Yu2018MultivalentBrushes,Sing2014ElectrostaticMorphology,Gelbart2007DNAInspiredElectrostatics}. \par
A classic example where the spatially varying ion correlation plays the dominant role is the vapor-liquid interface in ionic fluids. Since the pioneering experimental work by Buback and Franck\cite{Buback1972MeasurementsHalides}, it has been well recognized that ionic salts exhibit two-phase coexistence below a certain critical temperature\cite{Fisher1994TheCritiality,Schroer2012AFluids}. Analogous to vapor-liquid phase equilibrium in real gases, ionic salts dissolved in a solvent can also undergo phase separation into a high ion density `liquid' and a low density `vapor' phase because of ion correlations. This coexistence has also been predicted by theories, and molecular simulations, where both the phase boundary and criticality have been reasonably captured \cite{Fisher1993CriticalityBeyond,Fisher2005HowCriticality,Yan1999Hyper-parallelModel,Orkoulas1999PhaseEnsemble,Caillol2002CriticalRevisited}. Despite the progress in explaining the bulk thermodynamics, the interfacial behavior remain less addressed to our knowledge. Bresme and coworkers\cite{Gonzalez-Melchor2003SurfaceLiquids, Alejandre2009InterfacialLiquids, Gonzalez-Melchor2005MolecularLiquids} performed molecular simulations to study the interface of ionic fluids, including the density profile and surface tension. However, it is difficult to apply simulation methods away from the critical temperature due to the very low density of the vapor phase. On the theory side, two main methods have been invoked to model this vapor-liquid interface: non-local density functional theory\cite{Weiss1998RelevanceTheory,Groh1998Liquid-vaporFluid,Patrykiejew2011TheApproach} and the square-gradient theory\cite{TeloDaGama1980TheSalt,Weiss2000TheFluids,Lee1996DensityTheory}. These approaches describe the ion correlation in the inhomogeneous interfacial region using a functional form based on the bulk correlation function. In the density functional theory, the choice of the density to evaluate the local correlation is ad hoc\cite{Weiss1998RelevanceTheory,Evans1992DensityFluids}, which prevents its generalization to different systems. Whereas in the square-gradient approach, truncating the expansion of the free energy at the square-gradient level limits the applicability of this method only to systems where the concentration deviation from the bulk is small. The theory thus cannot be used to describe interfaces away from the critical point. Furthermore, none of the existing theories have been applied to systems where the cations and anions are not symmetric, either in terms of valency or ionic size. The asymmetry leads to an interphase electrostatic potential (Galvani potential) and local charge separation across the interface\cite{Fisher2005HowCriticality, Alejandre2009InterfacialLiquids}, which increases the complexity of both the physics and numerical solution. This gap in our understanding is a serious issue considering how ubiquitously asymmetric electrolytes exist. Therefore, a self-consistent theory and non-perturbative theory to describe the electrostatics at the vapor-liquid interface is necessary. \par

To accurately quantify the inhomogeneity of ion correlation is a great challenge majorly for the following two reasons. First, the correlation function needs to be resolved at two very different length scales, one associated with ion size (short-range) and the other with the interfacial thickness (long-range). Second, the ionic cloud is highly anisotropic on the two sides of the interface due to the huge difference in ion concentration between the two bulk phases. These two features have not been correctly captured in previous theoretical work because of the mathematical approximations involved. If only the local correlation is included, such as the WKB approximation widely used in double-layer theories\cite{Buff1963StatisticalProperties,Carnie2007TheLayer,Wang2015OnSurfaces}, it is in fact  impossible to generate a continuous interfacial profile. In this letter, we apply a modified Gaussian renormalized fluctuation theory\cite{Wang2010FluctuationEnergy,Agrawal2022ElectrostaticElectrolytes} to study the vapor-liquid interface for both symmetric and asymmetric ionic salts. A decomposition method is developed to decouple the short-range and long-range contributions in electrostatic correlation. Theoretical predictions are compared with simulation results from the literature. \par

We consider a system of an ionic salt with cations of valency $q_+$ and anions of valency $q_-$ dissolved in a solvent with dielectric function $\varepsilon ({\mathbf r})$. To accurately describe charge interaction, We assume that the ionic charge has a finite-spread given by distribution function $h_\pm(\mathbf{r}-\mathbf{r'})$ for an ion centered at $\mathbf{r'}$. We also include the excluded volumes of ions and solvent molecules.
The modified Gaussian renormalized fluctuation theory developed in our previous work\cite{Agrawal2022ElectrostaticElectrolytes} is used to model the bulk thermodynamics and the associated interface. This theory is based on a non-perturbative calculation of partition function using the Gibbs variational principle. In the absence of external fixed charge, the theory yields the following set of self-consistent equations for electrostatic potential, $\psi$, ion concentration, $c_\pm$, self energy $u_\pm$, and correlation function $G$ 
 \begin{eqnarray}
{-\nabla.[\epsilon(\mathbf{r})\nabla\psi(\mathbf{r})]}= q_{+}{c_{+}(\mathbf{r})} - q_{-}{c_{-}(\mathbf{r})} 
\label{eq:psi}
\end{eqnarray}
\begin{eqnarray}
{c_{\pm}(\mathbf{r})}= \frac{ \mathrm{e}^{\mu_{\pm}}}{v_{\pm}}\exp[\mp q_{\pm}\psi(\mathbf{r}) - u_{\pm}(\mathbf{r}) -v_{\pm} \eta(\mathbf{r})]
\label{eq:conc}
\end{eqnarray}
\begin{eqnarray}
{\textit{u}_{\pm}(\mathbf{r})}=\frac{q_{\pm}^2}{2}\int d\mathbf{r}'d\mathbf{r}'^{\prime}h_{\pm}(\mathbf{r'},\mathbf{r})G(\mathbf{r'},\mathbf{r'^{\prime}})h_{\pm}(\mathbf{r'^{\prime}},\mathbf{r})
\label{eq:selfe}
\end{eqnarray} 
\begin{eqnarray}
{-\nabla_{\mathbf{r}}.[\epsilon(\mathbf{r})\nabla_{\mathbf{r}}G(\mathbf{r},\mathbf{r'})]} + 2I(\mathbf{r})G(\mathbf{r},\mathbf{r'}) = \delta(\mathbf{r}-\mathbf{r'})
\label{eq:greens}
\end{eqnarray}
where $2I(\mathbf{r})= \epsilon(\mathbf{r})\kappa^2(\mathbf{r}) = c_{+}(\mathbf{r})q_{+}^2 + c_{-}(\mathbf{r})q_{-}^2$, $\mu_{\pm}$ are chemical potentials and $v_{\pm}$ are molecular volumes of the ions. $\epsilon({\mathbf{r}})=kT\varepsilon_{0}\varepsilon ({\mathbf r})/e^2$ is the scaled permittivity with $\varepsilon_{0}$ as the vacuum permittivity and $e$ as the elementary charge. $\eta(\mathbf{r})$ is the field accounting for the excluded volume effect, which can be expressed in terms of concentrations based on the incompressibility condition. The details of the theory are provided in Supplemental Material \cite{suppinfo}. \par

In the homogeneous bulk, where concentrations are uniform and $\epsilon(\mathbf{r})$ is a constant as $\epsilon_b$, the correlation function $G_b$ has a Debye-H\"{u}ckel style analytical form given by
\begin{equation}
G_{b}(\mathbf{r'},\mathbf{r''})  =  \frac{\mathrm{e}^{-\kappa_b|\mathbf{r'}-\mathbf{r''}|}}{4\pi\epsilon_b|\mathbf{r'}-\mathbf{r''}|}
\label{eq:gbulk}
\end{equation} 
For mathematical convenience we consider $h_\pm$ to be Gaussian, which leads to the following expression for the self energy $u_{\pm,b}$ 
\begin{equation}
u_{\pm,b} = \frac{q^2_\pm}{8\pi\epsilon_b} \left[\frac{1}{a_\pm}-  \kappa_b\exp\left(\frac{({a_\pm\kappa_b})^2}{\pi}\right)\operatorname{erfc}\left(\frac{{a_\pm\kappa_b}}{\sqrt{\pi}}\right)\right]
\label{eq:selfe_bulk}
\end{equation}
where $a_\pm$ is the radius of the ions associated with volume $v_\pm$. The first term on the r.h.s of Eq. \ref{eq:selfe_bulk} is the Born energy and the second term is the contribution from the ion correlation. By combining Eq. \ref{eq:conc} with Eq. \ref{eq:selfe_bulk}, chemical potentials can be expressed in terms of bulk concentration and electrostatic potential. The coexistence curve and the Galvani potential $\Delta \psi_G$ can thus be obtained by equalizing the chemical potentials and the grand free energy in the two bulk phases (see Sec. II of Supplemental Material \cite{suppinfo}).\par
To solve for the inhomogeneous interfacial region, spatially varying self-energy, $u_\pm(\mathbf{r})$, needs to be calculated. However, a full numerical solution for self energy will require discretizing Eq. \ref{eq:greens} at two length scales, one belonging to the ion size and the other to the interfacial width. This duality makes the exact numerical solution too expensive to be practically tractable. To overcome this barrier, the correlation function is decomposed into a short-range contribution $G_s$ and a long-range contribution $G_l$ as
\begin{eqnarray}
G(\mathbf{r'},\mathbf{r''}) = G_{s}(\mathbf{r'},\mathbf{r''}) + G_{l}(\mathbf{r'},\mathbf{r''})
\label{eq:greens_dec}
\end{eqnarray} 
$G_s$ is defined using the local dielectric permittivity $\epsilon(\mathbf{r})$ and local ionic strength $I(\mathbf{r})$ as
\begin{eqnarray}
-\epsilon(\mathbf{r}){\nabla_{\mathbf{r'}}^2G_s(\mathbf{r'},\mathbf{r'^{\prime}})} + 2I(\mathbf{r})G_s(\mathbf{r'},\mathbf{r'^{\prime}}) = \delta(\mathbf{r'},\mathbf{r'^{\prime}})
\label{eq:greens_short}
\end{eqnarray}
which has an analytical solution similar to Eq. \ref{eq:gbulk}: 
\begin{eqnarray}
G_{s}(\mathbf{r'},\mathbf{r'^{\prime}})  =  \frac{\mathrm{e}^{-\kappa(\mathbf{r})|\mathbf{r'}-\mathbf{r'^{\prime}}|}}{4\pi\epsilon(\mathbf{r})|\mathbf{r'}-\mathbf{r'^{\prime}}|}
\label{eq:gsol_short}
\end{eqnarray} 
Using the above expression of $G_s$, the short-range component of the self energy can be evaluated (i.e., $u_{\pm,s} = \frac{q_{\pm}^2}{2}\int _{\mathbf{r}',\mathbf{r}'^{\prime}}h_{\pm}G_sh_{\pm}$) to have \textcolor{red}the same functional form as Eq. \ref{eq:selfe_bulk} except that $\kappa_b$ and $\epsilon_b$ are replaced by $\kappa(\mathbf{r})$ and $\epsilon(\mathbf{r})$ (Eq. 24 in Supplemental Material\cite{suppinfo}). Next, the long-range contribution $G_l$ can be obtained by subtracting Eq. \ref{eq:greens_dec} from Eq. \ref{eq:greens}, which yields 
\begin{eqnarray}
{-\nabla_{\mathbf{r'}}.[\epsilon(\mathbf{r'})\nabla_{\mathbf{r'}}G_l(\mathbf{r'},\mathbf{r'^{\prime}})]} + 2I(\mathbf{r'})G_l(\mathbf{r'},\mathbf{r'^{\prime}})= S(\mathbf{r'},\mathbf{r'^{\prime}})\nonumber\\
\label{eq:greens_long}
\end{eqnarray}
where the non-local source term $S$ is
\begin{eqnarray}
S(\mathbf{r'},\mathbf{r'^{\prime}})  = \nabla_{\mathbf{r'}}.(\epsilon(\mathbf{r'})-\epsilon(\mathbf{r'^{\prime}}))\nabla_{\mathbf{r'}}G_s(\mathbf{r'},\mathbf{r'^{\prime}})\nonumber\\  - 2(I(\mathbf{r'}) - I(\mathbf{r'^{\prime}}))G_s(\mathbf{r'},\mathbf{r'^{\prime}}) \nonumber
\end{eqnarray}
It can be clearly seen from the above source term that $G_l$ accounts for all the non-local electrostatic effects associated with the spatially-varying ionic strength and dielectric permittivity, where the local values $I(\mathbf{r'^{\prime}}) $ and $\epsilon(\mathbf{r'^{\prime}})$ are taken as the reference. $S$ becomes zero as $\mathbf{r'}$ approaches $\mathbf{r'^{\prime}}$, which indicates that $G_l$ does not include any local contribution and is also free of the divergence problem.\par

To rationalize the aforementioned decomposition, it is important to look at the physical meaning of self energy. $u_\pm(\mathbf{r})$ is essentially the work needed to gather the constituting charges of the test ion from infinity to position $\mathbf{r}$ in the presence of a given electrostatic environment. This work depends on the charge spread of the ion $h_\pm$ as shown in Eq. \ref{eq:selfe}. The charge spread on the ion will only be crucial to the electrostatic forces originating in the close neighborhood of the test ion. In contrast, the electrostatic effects acting from far away are insensitive to the details of the charge distribution; or equivalently, the test ion can be taken as a point charge, $h_\pm(\mathbf{r}-\mathbf{r'}) =\delta_\pm(\mathbf{r}-\mathbf{r'})$. The nature of our decomposition of $G$ is such that $G_s$ only accounts for the local electrostatic environment, whereas all the long-range electrostatic effects are attributed to $G_l$. For most electrolyte solutions, ion radius is the smallest length scale in the system, while the long-range effects generally have characteristic length scales much larger than ion size. Therefore, Eq. \ref{eq:selfe} can be simplified by taking the same point limit of $G_l$ as
\begin{eqnarray}
\textit{u}_{\pm}(\mathbf{r})= u_{\pm,s}(\mathbf{r}) + \frac{q_{\pm}^2}{2}G_l(\mathbf{r},\mathbf{r})
\label{eq:selfe_h2}
\end{eqnarray} \par

Through the decomposition of $G$, we have successfully decoupled the short-range and long-range components of the electrostatic correlation. By taking the point-charge limit in Eq. \ref{eq:selfe_h2}, now we only need to discretize Eq. \ref{eq:greens_long} at a single length scale associated with the interfacial width, thus significantly reducing the numerical complexity. In the case of the planar geometry, such as the vapor-liquid interface, the problem is further simplified by performing a two-dimensional Fourier transform on $G_l$ in the plane parallel to the interface\cite{Xu2014SolvingEquations}. The details of the transform and the numerical scheme are presented in Sec. III in Supplemental Material\cite{suppinfo}. The decomposition of $G$ introduced in the current work has a similar physical basis as the Ewald summation commonly used in computing electrostatic interactions in molecular simulations\cite{Ewald1921DieGitterpotentiale, Toukmaji1996EwaldSurvey}. \par

\begin{figure}
 \includegraphics[width=\columnwidth]{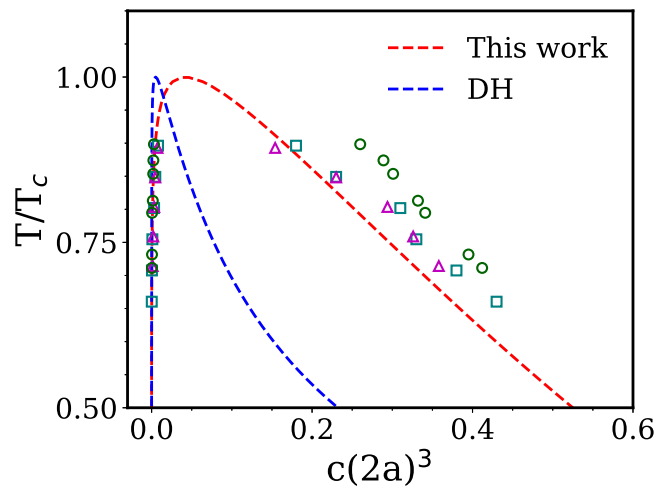}
  \caption{Phase diagram of liquid-vapor coexistence in symmetric ionic fluids plotted using reduced temperature ($T/T_\mathrm{c}$) and scaled total concentration of ions $c(2a)^3$. The lines represent theoretical predictions in comparison with the simulation data from Gonz\'{a}lez-Melchor et al.\cite{Gonzalez-Melchor2003SurfaceLiquids}(circles), Orkoulas et al.\cite{Orkoulas1994FreeSimulations} (squares), and Caillol et al.\cite{Caillol1997ACriticality} (triangles).}
\label{fig:binodal_symm}
\vspace{-1.5em}
\end{figure}

In the current work, we study the effect of inhomogeneous ion correlation on the structure and properties of the vapor-liquid interface for the case of constant dielectric permittivity, independent of the concentration. A constant $\varepsilon$ also facilitates the quantitative comparison with molecular simulations performed using the primitive model. We start with the bulk thermodynamics of symmetric salt, where $q_+ =q_- = 2$ and $a_+ = a_- = a$. Figure \ref{fig:binodal_symm} plots the phase diagram of the vapor-liquid equilibrium in terms of reduced temperature $T/T_\mathrm{c}$ where $T_\mathrm{c}$ is the critical temperature. By accounting for the finite charge spread on the ion, our theory predicts a much broader coexistence envelope compared to the point-charge Debye-H\"{u}ckel (DH) theory and is in quantitative agreement with three independent sets of simulation data\cite{Caillol1997ACriticality, Orkoulas1999PhaseEnsemble, Gonzalez-Melchor2003SurfaceLiquids}. In the homogeneous bulk, the self energy only contains the short-range component $u_{\pm,s}$, which depends on the details of the ion. Our results highlight the necessity of including the finite charge spread to accurately capture the short-range correlation. The short-range correlation thus also becomes a prerequisite to modeling the correlations in the interface. It should be noted that in Figure \ref{fig:binodal_symm}, the comparison is made at the same distance from the critical point. It is well recognized that fluctuation theories at the Gaussian level cannot accurately capture the critical point, a feature that is only possible to be reproduced through renormalization-group calculations\cite{Fisher1993CriticalityBeyond, Schroer2012AFluids}.

\begin{figure*}
\captionsetup[subfigure]{labelformat=empty}
\begin{subfigure}{\columnwidth}
 \includegraphics[width=\columnwidth]{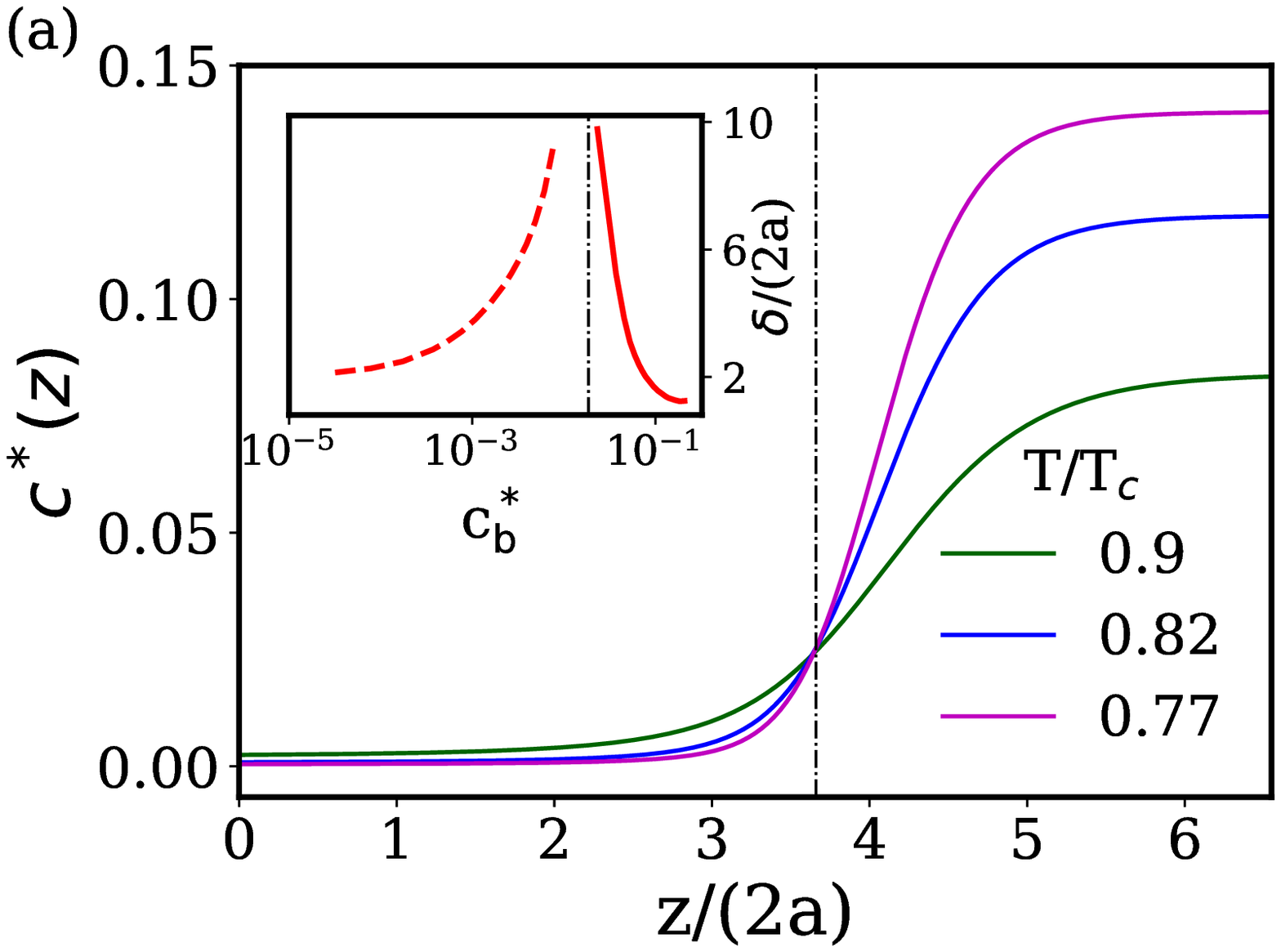}
        \caption{}
        \label{fig:concns}
    \end{subfigure}
    \hfill
\begin{subfigure}{\columnwidth}
 \includegraphics[width=\columnwidth]{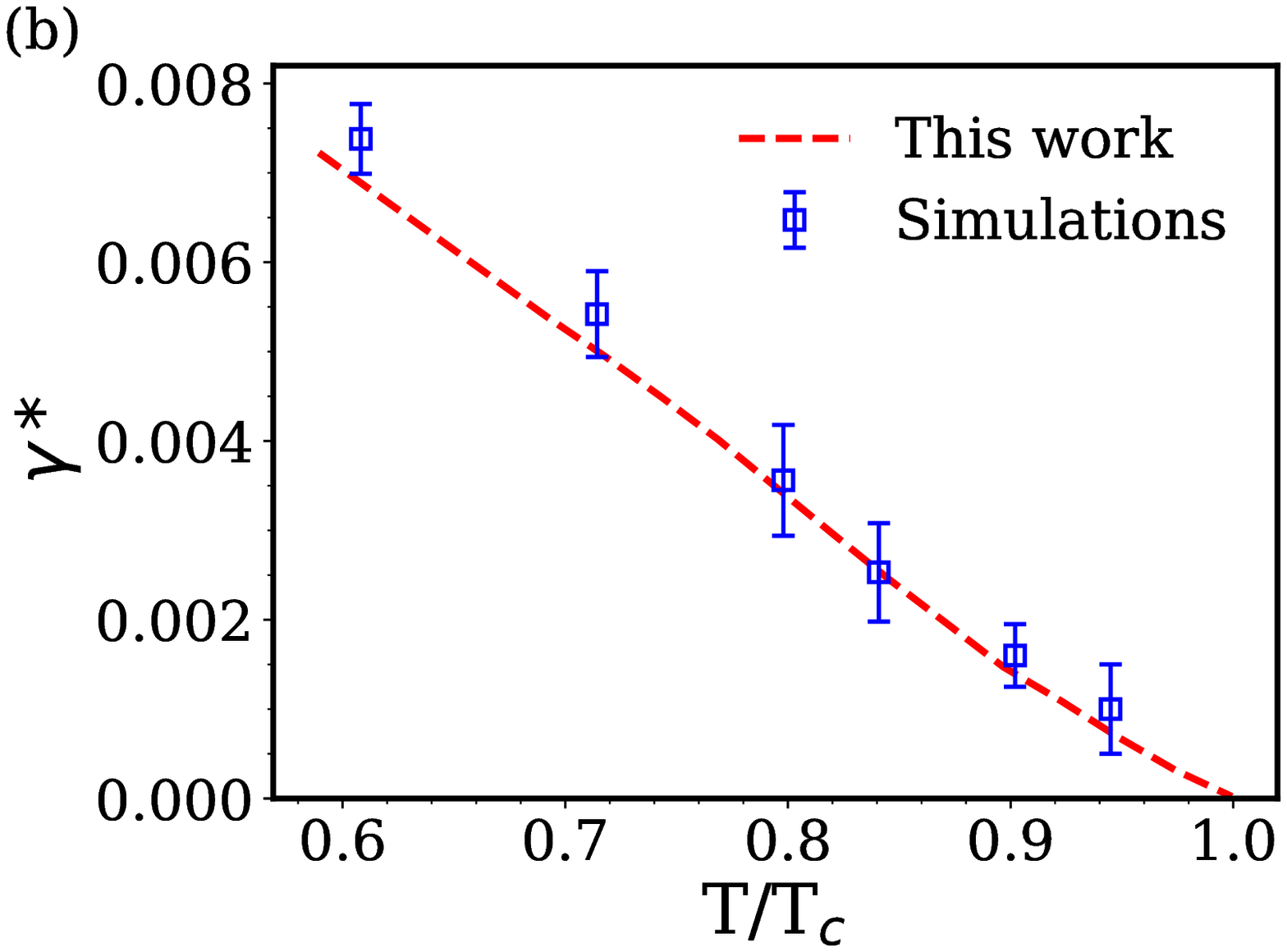}
        \caption{}
        \label{fig:gammas}
    \end{subfigure}
    \vspace{-1.5em}
\caption{Interfacial structure and properties for a symmetric salt, $\epsilon = 80$. a) Ion concentration $c^* = c(2a)^3$ profiles for different $T/T_\mathrm{c}$ with vertical dashed line denoting the Gibbs dividing surface. The inset plots the interfacial width on the vapor (dashed) and liquid side (solid) of the interface against the ion concentration in the corresponding bulk ($c^*_b$). b) Non-dimensional surface tension $\gamma^* =\gamma4\pi\varepsilon (2a)^3/(q_+q_-e^2)$ as a function of the reduced temperature $T/T_\mathrm{c}$ predicted by our theory in comparison with simulations of Gonz\'{a}lez-Melchor et al.\cite{Gonzalez-Melchor2005MolecularLiquids}}.
\label{fig:symm}
\end{figure*}
\begin{figure*}[ht]
\captionsetup[subfigure]{labelformat=empty}
\begin{subfigure}{0.67\columnwidth}
 \includegraphics[width=\columnwidth]{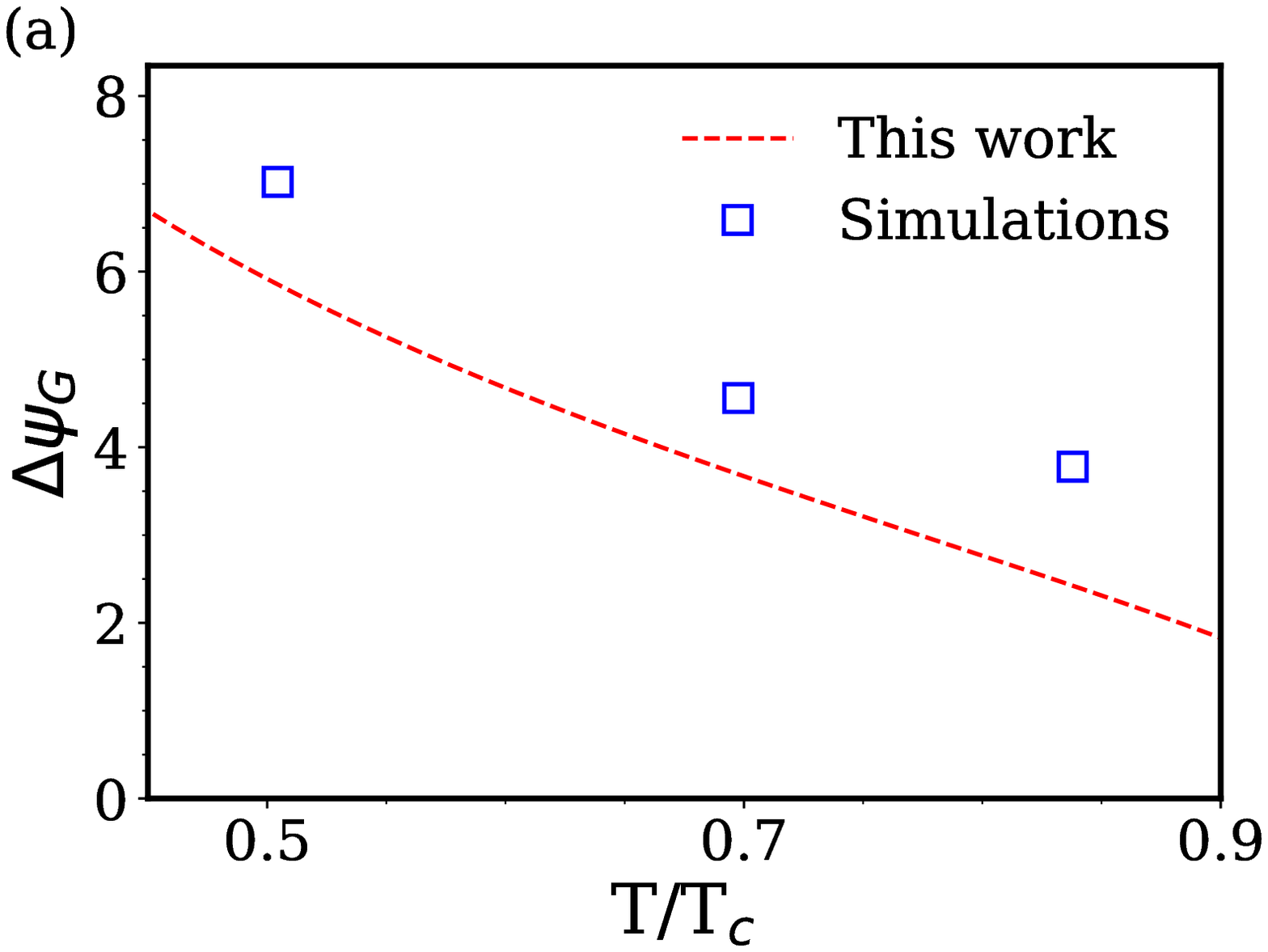}
        \caption{}
        \label{fig:galvani}
    \end{subfigure} 
\begin{subfigure}{0.67\columnwidth}
 \includegraphics[width=\columnwidth]{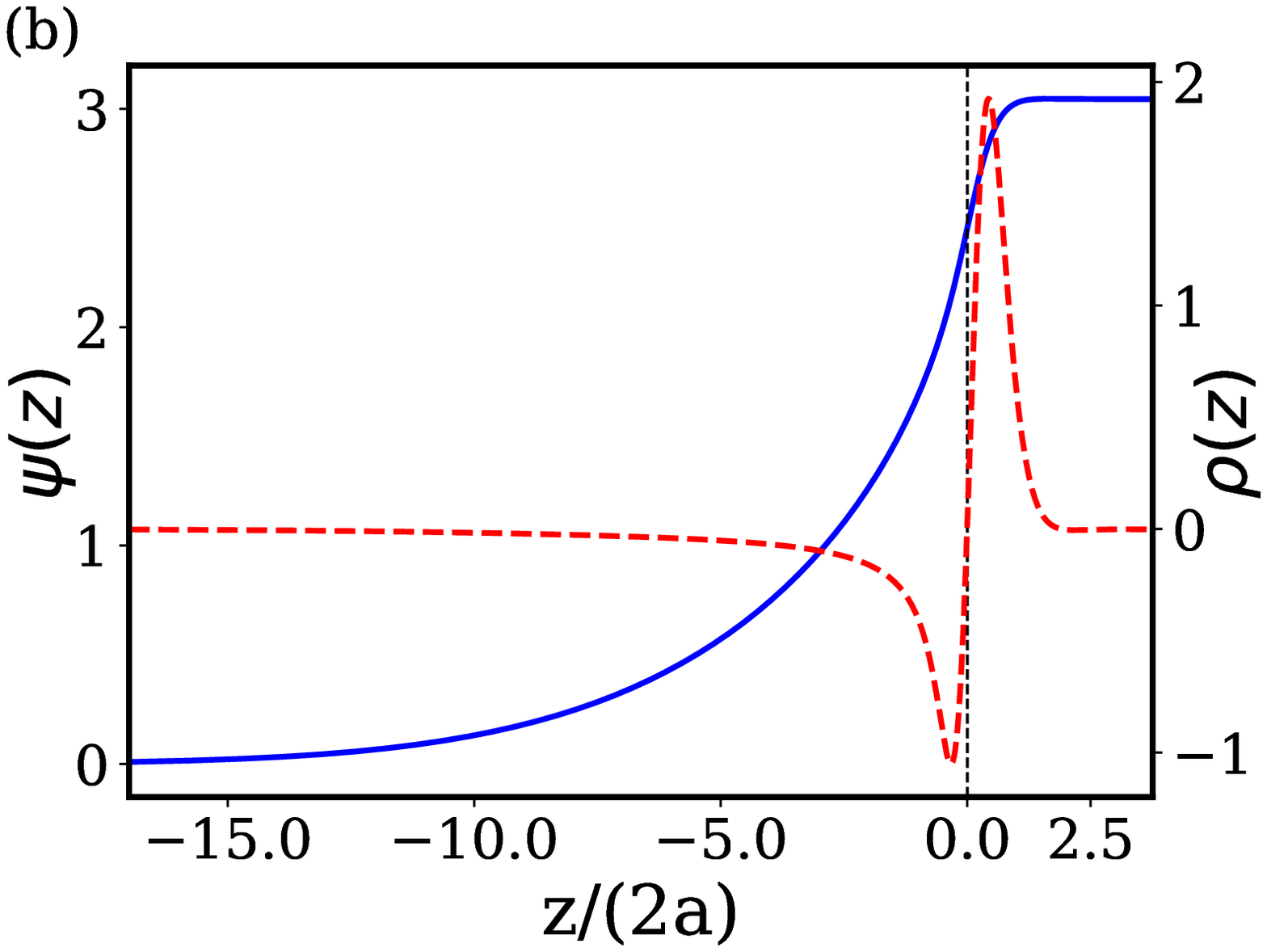}
        \caption{}
        \label{fig:psi}
    \end{subfigure}
\begin{subfigure}{0.67\columnwidth}
 \includegraphics[width=\columnwidth]{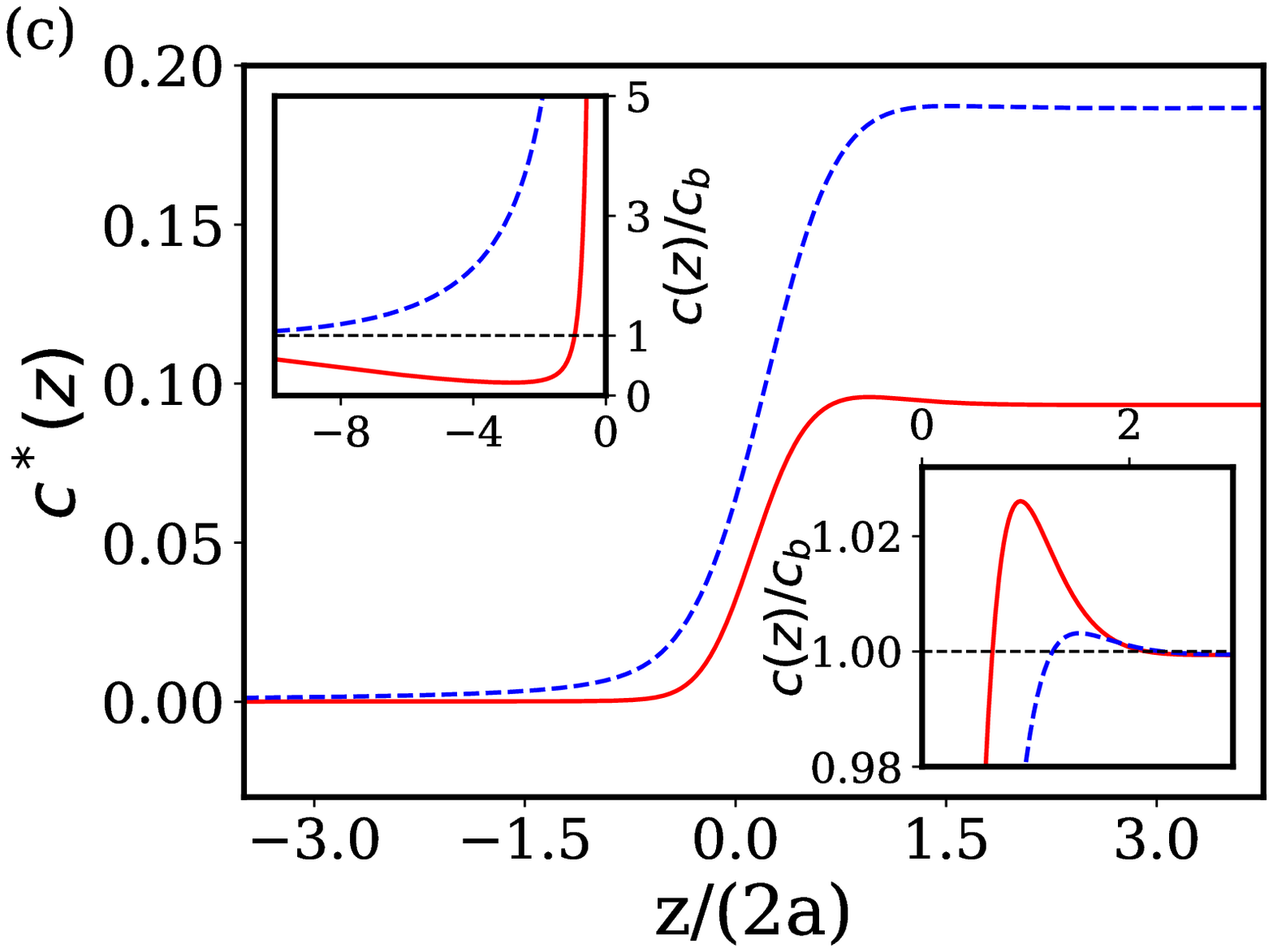}
        \caption{}
        \label{fig:concn_asymm}
    \end{subfigure}
\vspace{-1.5em}
\caption{Interfacial structure for a 2:1 asymmetric salt a) Theoretical predictions of Galvani potential $\Delta\psi_G$ in comparison with simulation data from Ref. \cite{Alejandre2009InterfacialLiquids} b) Electrostatic potential profile $\psi(z)$ (solid lines) and net charge distribution $\rho(z)$ (dashed lines). The vertical dotted line denotes the point of zero charge. c) Cation (solid) and anion (dashed) concentration profiles, $c^* = c(2a)^3$. The insets enlarge the double layer structure on the two sides of the interface, ion concentrations are plotted relative to their respective bulk values. $T/T_c = 0.76 $ and $\epsilon = 80$ for b) and c).}
\label{fig:asymm}
\end{figure*}

By resolving the correlation function at both short and long ranges, we can calculate the interfacial ion concentration profile between the two coexisting bulk phases. The concentration profiles for different values of $T/T_\mathrm{c}$ are shown in Figure \ref{fig:concns}. The profiles are shifted to have the same position of the Gibbs dividing surface (GDS). Including the long-range correlation effect is essential for capturing the continuous change in concentration from vapor to liquid. This diffused interface cannot even be created by theories that only include the local correlation. The importance of the long-range effect is also illustrated by the interfacial width on the two sides of the GDS, as shown in the inset of Figure \ref{fig:concns}. As the critical temperature is approached, the interfacial width on both sides increases and eventually diverges at the critical concentration, which is consistent with the divergence of the correlation length at the critical point as predicted by Lee and Fisher\cite{Lee1996DensityTheory}. Furthermore, the concentration dependence of the interfacial width on vapor-side $\delta_v$ is counter-intuitive. As the concentration in the vapor phase increases, its bulk correlation length (analogous to Debye screening length) decreases, which is expected to shorten $\delta_v$. The increase of $\delta_v$ predicted by our theory is a result of the long-range correlation effect from the bulk liquid phase to the vapor side. The two sides of the interface interfere with each other due to this long-range effect.\par 

Our theory also quantitatively captures the interfacial properties of the vapor-liquid interface. Figure \ref{fig:gammas} shows remarkable agreement of interfacial tension between our calculations and the simulation results from Gonz\'{a}lez-Melchor et al.\cite{Gonzalez-Melchor2005MolecularLiquids}. The quantitative agreement of both the phase coexistence curve and the interfacial tension for a wide range of temperatures validates the ability of our theory to accurately model interfaces with large interfacial inhomogeneities. For the same vapor-liquid interface, non-local density functional approaches have been found to overestimate interfacial tension values by a factor of three\cite{Weiss1998RelevanceTheory}. Our method is also superior to the square gradient theory-based approaches, which are only valid close to the critical temperature. The non-perturbative nature of our method allows us to capture inhomogeneous correlation beyond the square gradient level and thus guarantees its applicability to a variety of interfacial systems with steep concentration gradients. \par 

The modified Poisson-Boltzmann form of our equations enables us to conveniently include electrostatic potential and ion correlations in a unified framework. For the case of asymmetric salts, where cation and anion have different valencies or ion sizes, the difference in their self energies leads to local charge separation and an electrostatic potential profile across the interface. Here, we provide the first theoretical prediction of interfacial structure for a 2:1 salt. Figure \ref{fig:galvani} shows the Galvani potential $\Delta \psi_G$ predicted by our theory, in good agreement with the simulation results reported in literature\cite{Alejandre2009InterfacialLiquids}. The electrostatic potential profile and the net charge distribution are plotted in Figure \ref{fig:psi}. As the potential changes from $0$ in the vapor phase to a finite value in the liquid phase, its curvature and hence the net charge is forced to change sign at some intermediate point. Net positive and negative charges accumulate on the liquid and vapor sides, respectively. The two sides of the interface behave as charged objects with equal and opposite charges, essentially acting as double layers to each other. This dual double layer structure can be more clearly seen in the ion distribution in Figure \ref{fig:concn_asymm}.  On the vapor side, the net charge is negative and the electrostatic potential and ion distribution are similar to electrolyte solutions in contact with a positively charged surface. The left inset shows the depletion of cations compared to the bulk value, whereas anions are enriched. On the other hand, local excess of cations over anions can be seen in the right inset, as expected for a double layer next to a negatively charged surface. Similar self-energy-induced charge separation can also be observed in other interfaces, such as two immiscible fluids\cite{Wang2011EffectsMixtures, Sing2013Interfacialstudy} or micro-phase separated block copolymers\cite{Hou2018SolvationCopolymers, Sing2014ElectrostaticMorphology}.\par 

In Figure \ref{fig:concn_asymm}, it is worth noting that concentrations of both cations and anions on the liquid side are larger than their corresponding bulk values. This prediction can be explained by the strong correlation effect in the liquid phase. Similar counter-intuitive enrichment of cations is not expected on the vapor side due to its low ion densities. The cooperative enrichment of both counterions and coions is critical to explaining the phenomena of charge inversion in electrical double layers\cite{Agrawal2022ElectrostaticElectrolytes}. The over-accumulation of multivalent counterions next to the charged surface is stabilized by the presence of an excess amount of coions.\par
 
To conclude, the modified Gaussian renormalized fluctuation theory and the decomposition method developed in this work represent essential improvements over existing methods to model the vapor-liquid interface. The correlation function is decoupled into a short-range contribution associated with the local electrostatic environment and a long-range contribution accounting for the spatially varying ionic strength and dielectric permittivity. For most electrical double layers, the double-layer length scale is much larger than the ion size due to the long-range nature of Coulombic interactions. This allows us to separate the two length scales.
The non-perturbative and self-consistent nature of the theory allows the description of bulk thermodynamics and interface in a wide range of temperatures. A finite charge spread on the ion is necessary to accurately describe the short-range correlation and hence the vapor-liquid coexistence curve. Including the long-range correlation effect is essential for generating a continuous concentration profile across the interface. The resulting interfacial tension predicted by our theory is in quantitative agreement with simulation data for symmetric salts. We also provide the first theoretical prediction of the interface for an asymmetric salt, where the difference in ion correlation between cations and anions leads to an electrostatic potential profile and local charge separation on both the vapor and liquid sides of the interface. The ion distribution profiles on each side of the interface resemble an electrical double layer next to a charged surface. Because of high ion densities in the double layer on the liquid side, a cooperative enrichment of both counterions and coions is predicted. Finally, our theory is derived using a field-theoretical approach, which makes it straightforward to incorporate into other field-theoretical formulations, such as self-consistent field theory (SCFT) for polymers or Poisson-Nernst-Planck equation (PNP) for electrokinetic flows. \par 

\begin{acknowledgments}
R. W. acknowledges the support from the University of California, Berkeley. N. R. A. would like to thank Yannick A.D. Omar for helpful discussions regarding the numerical calculation of the Green's function. This research used the computational resources provided by the Kenneth S. Pitzer center for theoretical chemistry at UC Berkeley and the Savio computational cluster resource provided by the Berkeley Research Computing program.
\end{acknowledgments}
\bibliographystyle{apsrev4-2.bst}
\bibliography{intletter}

\end{document}